# Chern Insulator in magnetic-doped two-dimensional semiconductors

Dinh Loc Duong[*]

Department of Physics and Astronomy and Frontier Institute for Research in Sensor Technologies (FIRST), University of Maine, Orono, ME 04469 USA

[*]email: dinh.duong@maine.edu

**We propose an approach to induce nontrivial bands with non-zero Chern numbers by utilizing strong spin-orbit coupling in transition metal dichalcogenides with dopants. We demonstrate that a doped state near the valence-band edge induces band inversion with the hybridized host band, leading to topologically non-trivial properties. Calculations for V-doped $WSe_2$ and $WS_2$ confirm this mechanism. The coexistence of magnetic order and nontrivial topology in these systems offers a promising platform for exploring the quantum anomalous Hall effect.**

In high magnetic fields, free electrons in two dimensions are confined to circular orbits, occupying discrete energy states called Landau levels [1,2]. Experimentally, electrons can flow without resistance along the edges of the sample, creating a step-like Hall voltage drop between them, known as the quantum Hall effect [3,4]. In the quantum Hall effect, an external magnetic field is required, which is impractical for many applications, leading to the need for edge states in the absence of a magnetic field, known as the quantum anomalous Hall effect [5,6]. The topological properties of the Landau level, the Chern number, were proposed to explain the physics behind the quantization of the Hall conductance in the quantum Hall effect [7–9]. This leads to the demonstration of the quantum Hall effect without a magnetic field, as in the Haldane model of electronic band structure [9,10], later known as the Chern insulator. Although the model is not realistic, this approach enables the use of electronic band-structure calculations to search for novel materials for the quantum anomalous Hall effect by investigating their topological properties. In the quantum anomalous Hall effect, the edge current without dissipation is the most interesting feature [10,11], which can be used for low-energy-consumption, high-speed electronics [12] as well as quantum technologies [13], such as Majorana edge modes [12,14–18].

Several systems have experimentally demonstrated the quantum anomalous Hall effect, including magnetically doped topological insulators [19–22], the intrinsic antiferromagnetic topological insulator $MnBi_2Te_4$ [23–27], rhombohedral multilayer graphene [28], and twisted 2D materials [29–31]. However, the magnetic order, including orbital magnetization, in all these systems appears only at very low temperatures, which seems to hinder the observation of quantum anomalies at higher temperatures. Some other materials with non-zero Chern bands are predicted to have higher Curie temperatures [32], including monolayer transition-metal trichogenides [33],

TiSOH [34], 2D $M_2X_2$ [35], and $FeTaX_2$ [36]. Developing new approaches to expand this material family is essential for discovering novel materials exhibiting the high-temperature quantum anomalous Hall effect.

In this work, we propose a novel strategy to induce a Chern insulator through doping by causing band inversion between the doping state and the hybridized host band (Fig. 1). A magnetic dopant creates a spin-polarized doping state (e.g., spin-up) whose energy lies within the bandwidth of the host semiconductor's non-spin-polarized electronic band structure (Fig. 1(a), left panel). This spin-up band strongly hybridizes with the host's spin-up electron band, while leaving the spin-down band unaffected. This results in a spin-polarized band with an energy mismatch between the spin-up and spin-down bands (Fig. 1(a), right panel). We further consider certain dopants that can introduce an additional unoccupied spin-up state near the top of the spin-up hybridized band, without involving spin-orbit coupling (Fig. 1(b)). The strong spin-orbit coupling of the host band, if present, can cause a band conversion between the doping state and the host band, leading to a nontrivial band gap (Fig. 1(b), right panel). The non-zero Chern band is therefore introduced.

To demonstrate this concept, we present the band structure of V-doped $WSe_2$. Our previous study indicates that the inversion occurs between the empty doping state and the occupied hybridized states, inducing a strong doping effect and carrier-mediated long-range magnetic order in V-doped $WSe_2$, even at very low doping concentrations [37–43]. To investigate the detailed effect of spin-orbit coupling, the electronic band structure of V-doped $WSe_2$ was calculated with different SOC strengths (Fig. 2) [44]. Without SOC, the empty doping state is located on top of the hybridized band. Increasing SOC strength leads to band inversion, resulting in hybridization between the two bands (e.g., at 25% of full SOC strength). A stronger SOC keeps the doping state outside the bandwidth of the hybridized band, so there is no interaction between the two bands.

To investigate the topological characteristics of these two bands for 25% SOC strength, we extract tight-binding models for them using the Wannier90 [45] and PythTB codes [46]. Figure 3(a) shows the equivalent toy model that reproduces the band structure from the density functional theory calculation [47]. It consists of two orbitals located closely in a unit cell with a large lattice constant, revealing the doping characteristics of these two bands. The projection of the band structure onto one of the orbitals reveals a different color, indicating a mixed contribution from two orbitals within the bands, providing strong evidence for the band inversion (Fig. 3(b)). The loop integration of the Berry connection along the $k_x$ direction at different $k_y$ points, which represents the evolution of the Wannier centers across the unit cell from $k_y=0$ to $k_y=1$, yields different values from one edge of the unit cell to the other (Fig. 3(c)). Figure 3(d) shows the Berry flux through each plaquette, the smallest elementary square formed by four adjacent grid points, for the lower band. A clear high Berry flux appears at certain k-points, indicating a signature of the topological character of the band. The total summation returns the Chern numbers -1 and 1 for the lower and upper bands, respectively.

The above analysis demonstrates the Chern insulator characteristics of the crossing bands between the doping and hybridized bands at an SOC strength of 25%. However, at full SOC strength, these two bands do not cross. Therefore, tuning the doping concentration can be a strategy for inducing band crossing. We examine the band structure of V-doped $WSe_2$ at various doping levels by changing the supercell size. At lower concentrations (e.g., one V atom in a 12x12 supercell), the hybridized band is localized, without any band crossing. By increasing the doping concentration to one V atom in an 8x8 supercell (Fig. 4(b)), the hybridized band is more dispersed but not broad enough to cross the doped states. In contrast, at a higher concentration (e.g., one V atom in a 6x6 supercell), the hybridized state is more delocalized, leading to band crossings with the single-doping state (Fig. 4(c)). The analysis of the Bery phase (Fig. 4d) shows characteristics similar to those at 25% SOC strength, yielding Chern numbers of -1 and 1 for the lower and upper bands, respectively.

We further examine the band structure of V-doped $WS_2$ as another example of this strategy [48]. Unlike in V-doped $WSe_2$, the unoccupied doping state in V-doped $WS_2$ is always located at the top of the hybridized band, even in the presence of SOC. This band is quite localized. To form the band crossing between these localized states and the hybridized band, we consider the 8x8 supercell of 2 V atoms (Fig. 5(a)). The main idea in this case is to use the strong Coulomb interaction between V atoms to locate the doping state. Figure 5(b) shows the band structure of V-doped $WS_2$ with 2 V atoms. At a distance of approximately 3 lattice constants (~9.56 Å), the interaction between localized doping states is small, resulting in two slightly separated bands. To induce stronger interaction between these doping bands, two V atoms were brought closer together, at ~6.38 Å (Fig. 5(c)). As a result, one band is pushed down toward the valence band, and another is pushed upward within the band gap due to Coulomb interactions between the two localized bands. The lower-doping state is crossed by the hybridized band and induces a non-zero Chern-number band (Fig. 5(d)). Figure 5(e) shows the electronic structure of a ribbon cut from the periodic plane of the model with a width of 15 unit cells. It is obvious to observe the presence of the edge states of a nanoribbon located at the positions 0 and 15 (corresponding to two edges). The hybrid Wannier function center also reveals the discontinuity at the edge.

*Discussions*: While magnetic order is important in our proposed model, understanding how it forms in these materials remains a challenge. In all our calculations, the ferromagnetic state is initialized. Experimentally, both V-doped $WSe_2$ and $WS_2$ revealed ferromagnetic order at room temperature. The mechanism for establishing long-range ferromagnetic order in V-doped $WSe_2$ has been studied extensively, confirming free-carrier-mediated exchange coupling [37,39,40]. In contrast, the mechanism of V-doped $WS_2$ remains unclear. While the magnetic order can be formed in V-doped $WSe_2$ with a doping concentration below 0.5%, a larger doping concentration is required in V-doped $WS_2$. It may reflect the electronic band structure of V-doped $WSe_2$ and V-doped $WS_2$: one is a merging band or carrier-mediated, and the other is mediated by localized doping states or impurities band. Increasing doping concentration can induce band merging in V-doped $WS_2$. Increasing doping concentration is the key approach to inducing a non-zero Chern band. Nevertheless, the magnetic order of V-doped $WSe_2$ has disappeared at high doping concentration [38]. In contrast, the magnetic order of V-doped $WS_2$ may be maintained. Therefore, we suggest that the search will be more possible for V-doped $WS_2$. We note that our analysis is based on periodic band-structure calculations, whereas dopants are randomly distributed in the host materials experimentally. Nonetheless, the nature of topological properties does not strictly apply to crystal materials [49–51]. A number of studies have proposed that topological order can arise in disordered systems [51–53]. Therefore, our proposal is also valuable for realizing topological disorder in materials.

In summary, we propose a possible topological property in a magnetic-doped semiconductor. The strategy is to utilize the electronic band structure of dopant states, hybridized with the host states, to induce band inversion, owing to the strong spin-orbit coupling in two-dimensional semiconductors. We use two examples of recent experimental developments in ferromagnetic semiconductors, V-doped WSe$_2$ and WS$_2$, to demonstrate the concept. This provides a platform for searching for novel Chern insulators with high transition Curie temperatures, as well as for disordered magnetic topological materials.

**Acknowledgements**: This work was supported by a start-up grant from the University of Maine.

**Data availability**: The data that support the findings of this article are publicly available upon publication. The data are available from the authors upon reasonable request.

**FIGURES AND FIGURE CAPTIONS**

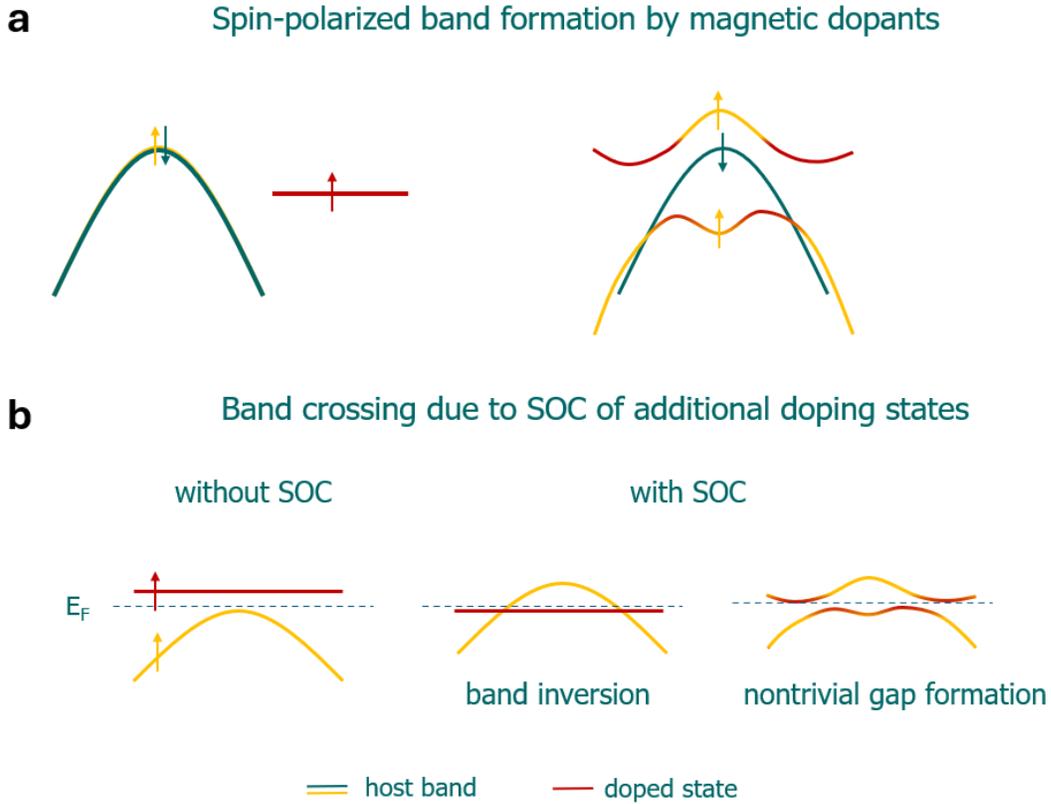

FIG. 1. Schematic of the formation of spin-polarized band crossing. (a) Formation of a hybridized band between the dopant and host atoms. (b) Band inversion by additional doping states located on the top of the hybridized band.

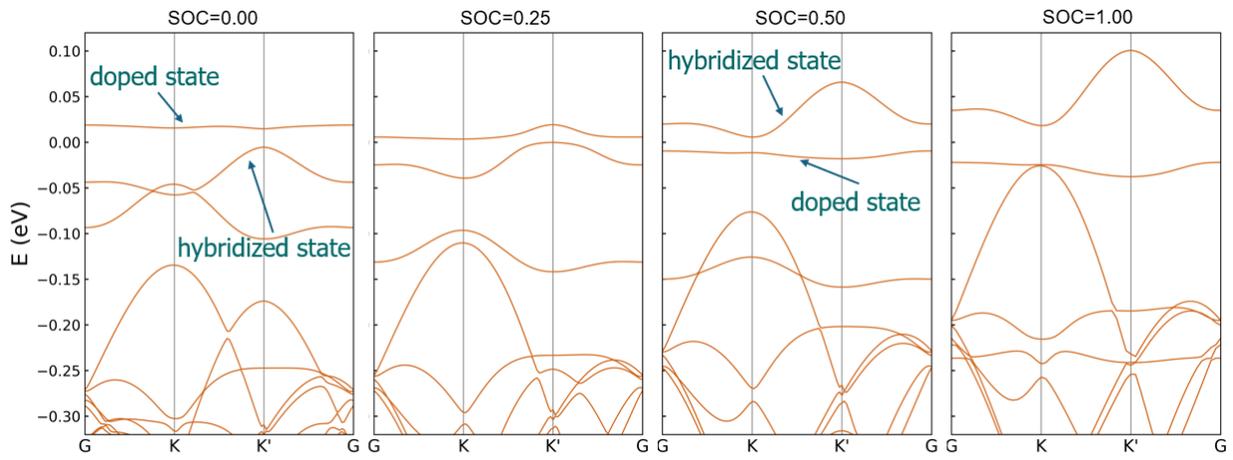

FIG. 2. Electronic band structure of V-doped WSe2 (8x8 supercell) with different strengths of spin-orbit coupling. Band crossing appears when the spin-orbit strength is about 25%.

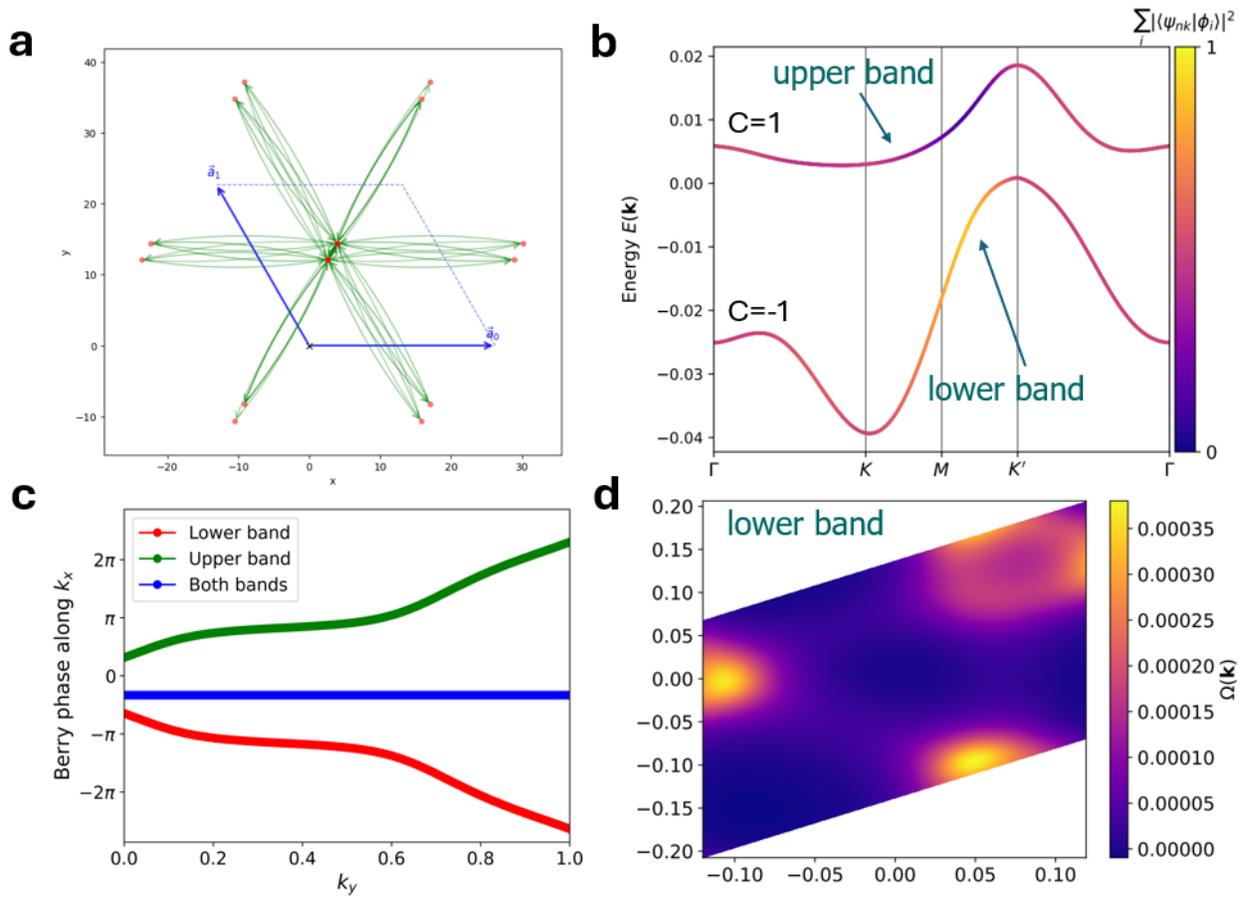

FIG. 3. Topological analysis of the band structure of V-doped WSe2 with 25% strength spin-orbit coupling. (a) Equivalent tight-binding model. (b) Projection of orbital contribution of two orbitals to lower and upper bands. (c) Berry phases around the BZ in the $k_x$ direction. (d) Berry flux through each plaquette for the lower band.

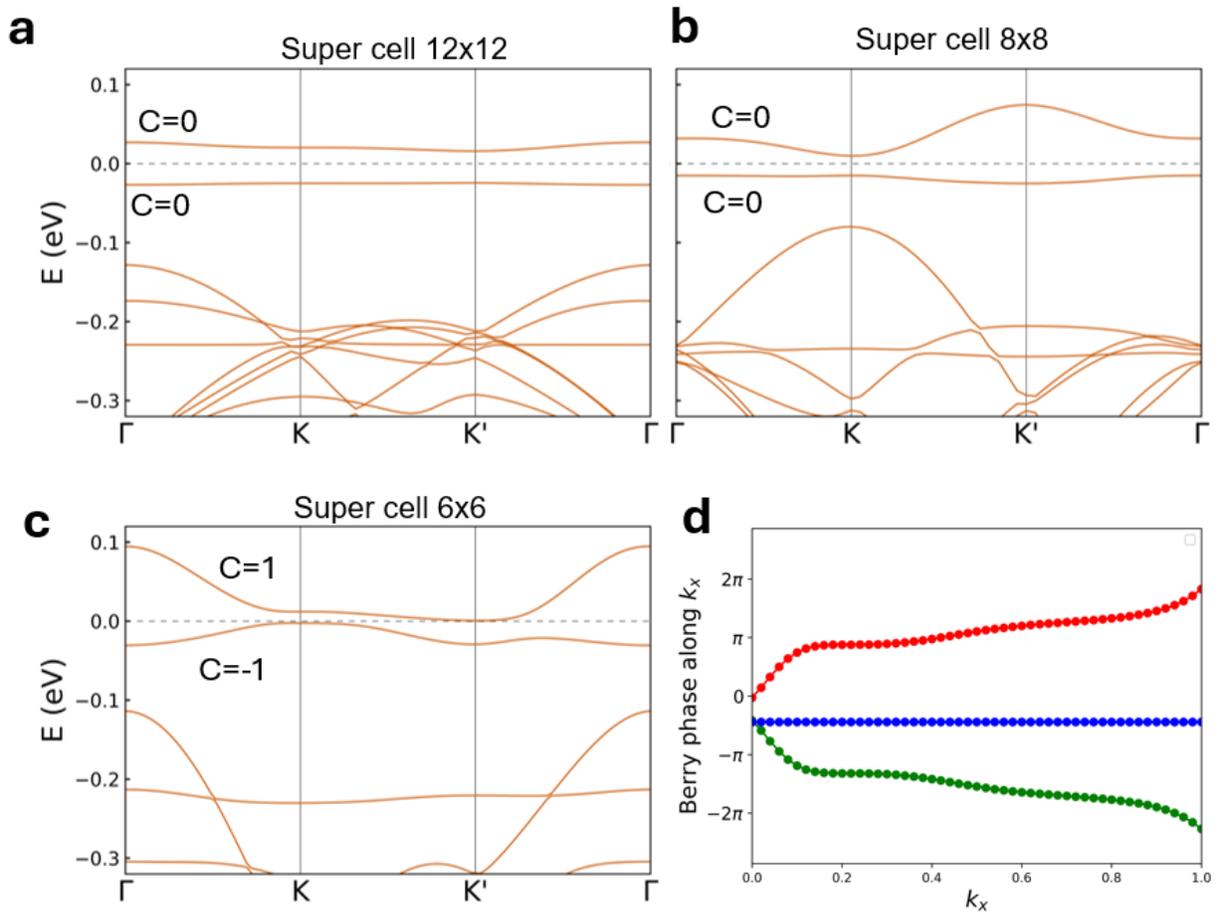

FIG. 4. Electronic band structure of V-doped WSe2 with full spin-orbit coupling strength at different doping concentrations. One atom is considered in a 12x12 (a), 8x8 (b), and 6x6 (c) supercell. (d) Berry phases around the BZ in the $k_x$ direction for the case of electronic band structure in (c).

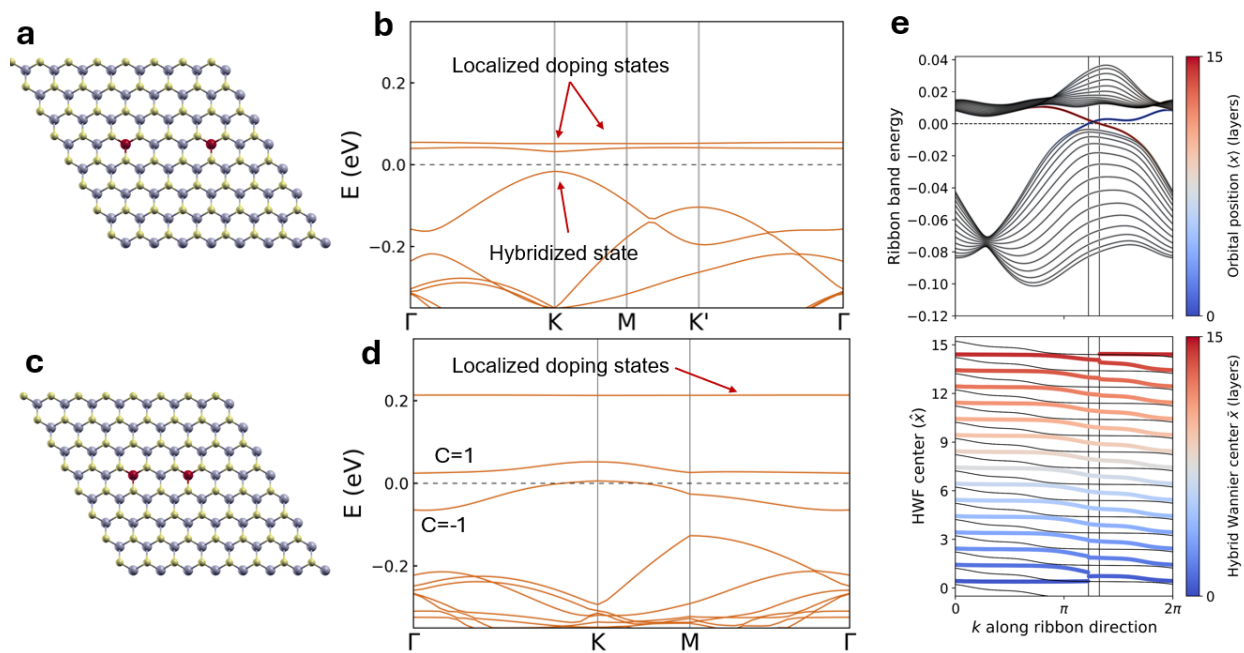

FIG. 5. Atomic structure of V-doped WSe$_2$ with two atoms in an 8x8 supercell and its electronic band structure. Two atoms are far (a,b) and close (c,d) together. (e) Edge-state analysis of a nanoribbon with a width of 15 unit cells.